\documentstyle[eqsecnum,aps]{revtex}
\input epsf.sty
\newcommand{\ket}[1]{| #1 \rangle}

\begin{document}

\draft


\title{Quantum Information is Physical
\footnote{To appear in
Superlattices and Microstructures. Special Issue
on the occasion of Rolf Landauer's 70th Birthday, ed. S. Datta.
See cond-mat/9710259.}}

\author{David P. DiVincenzo}
\address{
IBM Research Division,
Thomas J. Watson Research Center,
P. O. Box 218,
Yorktown Heights, NY 10598 USA
}
\author{Daniel Loss}
\address{
Department of Physics and Astronomy,
University of Basel,
Klingelbergstrasse 82,
4056 Basel, Switzerland
}

\date{October 23, 1997}

\maketitle

\begin{abstract}
We discuss a few current developments in the use of quantum
mechanically coherent systems for information processing.  In each of
these developments, Rolf Landauer has played a crucial role in nudging
us and other workers in the field into asking the right questions,
some of which we have been lucky enough to answer.  A general overview
of the key ideas of quantum error correction is given.  We discuss how
quantum entanglement is the key to protecting quantum states from
decoherence in a manner which, in a theoretical sense, is as effective
as the protection of digital data from bit noise.  We also discuss
five general criteria which must be satisfied to implement a quantum
computer in the laboratory, and we illustrate the application of these
criteria by discussing our ideas for creating a quantum computer out
of the spin states of coupled quantum dots.
\end{abstract}

\bigskip

\section{Personal note on Rolf Landauer}

We are extremely pleased to be able to add our contribution to this
collection of works, by many eminent authors in a wide spectrum of
fields, in honor of Rolf Landauer's lifetime of contributions to our
understanding of the physical world.  We will say a lot more about
what those contributions have meant for us below, but we might note
one important motivation which we have been given by Rolf's many
battles with the producers of conference books and special volumes
such as this one!  Rolf believes strongly that the written word should
be taken seriously, and that contributions, seriously written, should
not languish on an editor's shelf waiting for slaggart authors or
testy co-editors to do their neglected duty.  So we have, first and
foremost, been assiduous in delivering this contribution to
Prof. Datta by the announced deadline!

As will be evident in our technical discussion below, we are largely
mathematical physicists.  We would say that Rolf generally takes a
jaundiced view of such creatures; for him, there is an absolute need
for {\em explanation} and {\em understanding} of scientific truths in
the full human sense, not in the purely formal and mathematical sense
with which those of our species often content themselves.  Rolf has
challenged us throughout our careers, from our Green function studies
(as mere children) of conductance fluctuations and Aharanov-Bohm
effects in mesoscopic conductors, to our very recent brazen assertions
about the efficacy of error correction techniques in quantum
information processing---he has challenged us to explain, with as much
clarity and insight as we could muster, the basic underlying reasons
why the assertions that we were making were really true.  And, as we
feel we have come to understand better as we struggle towards our own
scientific maturity, it is in the answers to these questions, posed
relentlessly to everyone (we suspect) whose article appears in this
volume, and not the successful summations of a diagram series, from
which true, significant scientific progress is produced.  So, we thank
Rolf greatly for these promptings which he has given to our own work.

\section{Quantum error correction}

As one specific illustration of the remarkable flowering of scientific
progress which has resulted from the Landauer way of thinking, we
would like to tell the story, in which one of us (DDV) has been a
player, of the development of the theory of quantum error correction.
This theoretical area, which today has many ardent practitioners (some
more about them below), is remarkable in that it was believed to be a
strictly non-existent subject as little as two years ago (at this
writing, September 1997).

Rolf contributed to the belief that quantum error correction could not
be done\cite{Lan1,Lan2}, as indeed it could not in the original
conceptions of quantum computing.  His message seemed to be
compelling--- quantum information is a form of analog information.  In
many respects this statement is entirely correct: what we mean by a
qubit is a two-state quantum system which, unlike the conventional
bit, can be a continuum of possible states as specified by its
wavefunction:
\begin{equation}
\Psi=\alpha|0\rangle+\beta|1\rangle.
\end{equation}
Here $\alpha$ and $\beta$ are arbitrary complex numbers, apart from
the normalization condition.  (Actually, the permissible state of a
qubit is more general; it can be in a mixed state described by a
density operator.)  The availability of this continuum of states is
intrinsic to quantum computation; it appears that any attempt to
restrict the qubit to one of a discrete set of states will nullify the
great potential capabilities of quantum algorithms.  It would make no
physical sense to do so in any case, since the unitary evolution of
the quantum state under the action of a Hamiltonian is naturally a
continuous-time process.

Noting this feature in the earliest speculations about quantum
computation, Landauer challenged the workers in this infant community
with the statement that quantum computation could not be error
corrected, and thus lacked a crucial element that defines computation
itself.  He based these criticisms on the well known defect of
classical analog computing:  Since all states of such a device, and of
the quantum computer, are legal computational states, it is argued
that there would be no way to distinguish a state to which some noise
had been added.  Thus, there is no correction mechanism, and the
computation immediately starts to drift off track.  The imagery which
has been used by Landauer of the situation in digital computation is
that of a particle moving along a track of the ``standard'' digital
state, with very high potential-energy walls continually
``restandardizing'' the state as the computation proceeded (by the
movement of the ball in this potential-energy maze).

Rolf quite correctly saw no hint of restandardization in any of the
quantum computer implementations which were initially discussed, and
he offered up detailed criticisms of several of the schemes; for
instance\cite{Lan1}, in a computation scheme proposed by Benioff in
which the computation advanced by the propagation of a wavepacket in a
one-dimensional periodic potential, Landauer pointed out that the
phenomenon of one-dimensional localization made it exponentially
likely that before completing, the wave packet/computation would be
turned around by localization and relection.

Thus Landauer's criticism hung as a bit of a pall over the earlier
days of quantum computation, sprinkling a little rain on the otherwise
cheerful and naive quantum computation parade.  But Landauer's
criticisms had an extremely important effect; it got a couple of very
original minds like Peter Shor and Andrew Steane to {\it think} about
whether restandardization {\it could}, despite appearances, be
performed in quantum computation.  (Berthiaume {\em et al.}\cite{Bert}
and Bennett {\em et al.}\cite{Ben1,Ben1a} were also pursuing lines
which was not too far distant from what turned out to be the correct
one.)

Part of the discovery which Shor\cite{Shor}, Steane\cite{Steane}, and
then many others developed was a relatively obvious one, namely that
quantum states could be {\em encoded}.  In classical information
theory, coding just refers to the use of a string of bits to stand in
for the value of one bit (or perhaps a smaller block of bits).  The
idea is that the redundancy in this encoding allows errors (at least
some errors) to be caught and repaired; such is the standard practice
in digital communications.

It should, of course, be not at all obvious how redundancy can be of
any use in quantum computation.  Redundancy is apparently not very
useful in conventional analog computation; in addition, since quantum
states cannot be ``cloned'' or copied\cite{Woo,BBM}, it would seem
that even the simplest kind of redundancy is not even possible in
principle.  What Shor and Steane discovered was an ingenious way to
use an entirely different fundamental quantum property, {\em
entanglement}, in the service of redundancy and error correction.

Entanglement, introduced into quantum physics by
Schr\"odinger\cite{Sch1,Sch2} in 1935, is at one level a fairly
prosaic mathematical feature of the wavefunction of two (or more)
particles.  It refers to the fact that the composite wavefunction may
not be expressible as the product of the states of the two individual
particles.  For example, for two qubits one may have a state like
\begin{equation}
\Psi^-={1\over\sqrt{2}}(|01\rangle-|10\rangle)\ne\phi_A\times\phi_B.
\end{equation}
This inescapable feature of the fundamental principles of quantum
physics has a variety of consequences which, depending on your point
of view, are deep, profound, bizarre, ridiculous, or some combination
of the four.  (Indeed, Schrodinger recognized the peculiarity of all
this; we highly recommend his brilliantly written
articles\cite{Sch1,Sch2} from 1935-6 on the subject.)  One simple
consequence was touched on above: the state of an individual particle
cannot be described by a pure state $\phi_A$ in general; the density
operator is given by tracing the full state $|\Psi^-
\rangle\langle\Psi^-|$ over the other part of the quantum system.  It
may be said that all randomness in quantum physics, as described by
the probability amplitudes in the density operator, arises from
entanglement.

So, what Shor and Steane started with was the idea that the logical
$|0\rangle_L$ and $|1\rangle_L$ of a qubit could be coded as two
orthogonal entangled quantum states; the simple example which is
central to both of their analyses is the coding into the three-qubit
states
\begin{eqnarray}
|0\rangle_L=|000\rangle+|011\rangle+|101\rangle+|110\rangle,\\
|1\rangle_L=|111\rangle+|100\rangle+|010\rangle+|001\rangle.
\end{eqnarray}
(we will leave out normalization factors here and elsewhere.)  A few
elementary observations about this coding are in order: Since the two
coded states are orthogonal, this is in fact a good coding for the
entire qubit spanned by $|0\rangle$ and $|1\rangle$; that is,
$\alpha|0\rangle+ \beta|1\rangle$ is coded
as $\alpha|0\rangle_L+\beta|1\rangle_L$.  The second, more important
point has to do with the role of the entanglement of these states.  It
is easy to see that the density operator of any one of the three
qubits in the codeword is an equal mixture of $|0\rangle$ and
$|1\rangle$, whether the coded qubit is  in the $|0\rangle_L$ or
$|1\rangle_L$ state.  As Steane says it, the information about which
state the coded qubit lies in is not contained in any single one of
the coding qubits; it is spread out into a ``multi-particle
interference,'' which is set up as a result of the entanglement of the
encoded state.

One would like to think that this lack of information in any one
particle means that the coded qubit could be recovered after any
interaction by the environment with one of the three coded qubits,
since the way that a qubit gets disturbed is by a successful external
measurement of its state.  This turns out not to work for the
three-qubit example above, but it is very definitely on the right
track.  In fact, this reasoning leads us to what's wrong with the
three-qubit code; it is easy to see that, while a demon making
measurements on one qubit cannot learn whether the coded qubit is a
$|0\rangle_L$ or a $|1\rangle_L$, it can easily learn whether the
coded qubit is in the state $|0\rangle_L+|1\rangle_L$ or
$|0\rangle_L-|1\rangle_L$.  This is so because of the simple algebraic
fact that
\begin{equation}
|0\rangle_L\pm|1\rangle_L=|(0\pm 1)(0\pm 1)(0\pm 1)\rangle.
\end{equation}
Thus, the coding of these states involves product states rather than
entangled states, and is therefore quite ineffective at hiding the
state of the coded qubit from the environment which is ``measuring''
in this diagonal basis.

This result suggests that a quantum code which is completely effective
against single-qubit error must be able to make the information about
the coded state suitably recondite in {\em any} basis; one might
presume that this will work by making the state appropriately
entangled in any basis.  One might also have a brief worry that there
will some fundamental feature of quantum mechanics which makes the
right kind of entanglement unavailable.  Fortunately this is not the
case, and indeed there was no point in the history of the subject when
this worry held sway, as Shor and Steane both immediately found
solutions inspired by the three-qubit entangled states.  Their
discoveries, however, involved the invention of new forms of
multi-particle entangled states, in Shor's case a pair of nine-bit
entangled states, and in Steane's case a pair of seven-bit states.
Just in case the reader would like to contemplate these new sorts of
entangled states explicitly, we would like to write down a pair of
states which were discovered some time after by Laflamme {\em et
al.}\cite{Laf} and independently by us\cite{Ben1a}; they are
five-qubit states, and they are the smallest states for which one-bit
error correction is fully effective:
\begin{eqnarray}
\ket{0}_L &=& \ket{00000}\label{sym0}\\
&+& \ket{11000} +\ket{01100} +\ket{00110} +\ket{00011} +\ket{10001}
\nonumber \\
&-& \ket{10100} -\ket{01010} -\ket{00101} -\ket{10010} -\ket{01001}
\nonumber \\
&-& \ket{11110} -\ket{01111} -\ket{10111} -\ket{11011} -\ket{11101}
\nonumber
\end{eqnarray}
and
\begin{eqnarray}
\ket{1}_L &=& \ket{11111}\label{sym1}\\
&+& \ket{00111} +\ket{10011} +\ket{11001} +\ket{11100} +\ket{01110}
\nonumber \\
&-& \ket{01011} -\ket{10101} -\ket{11010} -\ket{01101} -\ket{10110}
\nonumber \\
&-& \ket{00001} -\ket{10000} -\ket{01000} -\ket{00100} -\ket{00010}
\nonumber .
\end{eqnarray}
These new states turn out not only to have the important features we
have mentioned for error correction, but they also all appear to
exhibit a type of quantum non-locality which is of great interest in
the foundations of quantum theory, and may indeed be viewed as the
natural extension of the work starting with the seminal paper of
Einstein, Podolsky and Rosen, and continuing to the discovery of the
Greenberger-Horne-Zeilinger states\cite{Per}.

A remarkable feature of the error-correcting properties of these
states is that, despite Landauer's and other's expectations, the
process of error correction has an essentially digital rather than
analog character\cite{DSS}.  We mean this in a rather straightforward
operational sense: we find (DiVincenzo and Shor\cite{DS} give the
details for the five-qubit code) that error detection involves the
performance of a series of binary-valued quantum measurements.  Then
these bit values provide an instruction for an error detection step,
which involves a discrete rotation of a specific one of the five qubit
states.  It seems that we can say that the reason for this essentially
digital character, despite the analog structure of the state space, is
that the code states arrange that any error which the environment can
cause by operating (in any arbitrary way) on a single qubit acts in a
subspace {\em orthogonal} to the state space of the coded qubit
itself.  Thus, the complex coefficients are, to very high accuracy,
completely untouched by the error process, and all the error detection
and correction steps can work in a way which is oblivious to their
values.  Thus, Landauer has turned out to be wrong on this point; but
by his challengings, he has opened up an entirely unanticipated line
of investigation in the fundamental properties of quantum mechanics.

We will not pursue the details here of how this story has further
developed, although we have told just the beginning of what has been a
tremendous development in the last two years: The formal conditions
have been constructed for entangled states to be effective as
correctable code states\cite{Ben1a,KL}, which in turn led to the
discovery of a powerful group-theoretic framework which permitted the
classification of essentially {\em all} interesting quantum codes of
arbitrary block size\cite{Got,CSSR}; it has been found how error
correction can be implemented fault tolerantly\cite{FT}, that is, in
such a way that it is insensitive to errors that occur during the
error detection operations themselves; finally, this has in turn led
to a discovery of protocols that permit fully general quantum
computation to be performed in the presence of errors and
decoherence\cite{Pres}.

All of this has amounted to a revolution in the way we theorists think
about the future prospects for quantum computation.  As of this
writing, we would say that Rolf, while being quite impressed by these
developments, and being convinced that they really make quantum
computing not such a hopeless enterprise as he once had thought, takes
a slightly jaundiced view of these developments.  He will make sure
that we don't loose sight of the fact that this theoretical
development emphatically does {\em not} solve all (or perhaps even
any) of the problems which stand in the way of making progress in the
laboratory today on the construction of a quantum computer.  We expect
that Rolf's wry commentaries on our efforts will continue to nudge
our mathematical efforts in the direction of real progress.

\section{The quest to embody quantum information physically}
\label{sec:two}

We would like to turn now to a different aspect of quantum information
and its physical embodiment, which also has its roots in the
promptings of Rolf Landauer.  Early in the development of the theory
of quantum computation, Benioff\cite{Beni}, Feynman\cite{Feyn}, and
Deutsch\cite{Deut} suggested, by way of existence proofs, some
abstract models of quantum systems whose dynamics would result in the
execution of some computation.  Landauer criticized this work very
pointedly; he emphasized that computers are made out of physical
apparatus and not out of Hamiltonians.  And he made the point, as we
discussed above, that it does not constitute a serious model of
computation if the imperfections in these apparatuses are not dealt
with in the analysis--he showed how several of the abstract systems
which had been proposed, if taken seriously, would exhibit very severe
flaws which would preclude their being a serious physical basis for
computation.

So the situation sat in around 1994, when Shor revolutionized the
field\cite{Shorbig} and raised to an altogether higher level of
significance the question of whether quantum information really could
be embodied physically---if a quantum computer to do factoring really
could be built.  We believe that the existing commentaries which had
been put in the literature by Rolf made the exploration of possible
physical embodiments a much soberer and realistic undertaking than it
would have been.  Models which were current at that time included the
spin-polymer concept of Lloyd\cite{Lloy}, and the atomic force
microscope ``clockwork'' computer of DiVincenzo\cite{crock}.  Both
were motivated by the Landauer criterion of proposing actual physical
embodiments of instruments which could manipulate quantum information,
and which were thus subjectable to realistic criticisms based on the
criteria of experimental physics.

However, both the Lloyd and DiVincenzo efforts look, at the remove of
three years, very naive and incomplete compared with the much more
impressive recent proposals to realize quantum computation in the
laboratory.  One enormously impressive stream of ideas and proposals
have flowed forth from the group of Zoller and Cirac\cite{CZ}.  These
workers, armed with a very deep understanding of the state of the art
in atomic physics and quantum optics, and informed by the very
perceptive general formulations of quantum information processing as
conditional dynamics produced by Artur Ekert at a very important
conference in 1994\cite{Eck}, have, we think, largely passed the
Landauer test of giving a thoroughly complete, analyzable, and
testable scenario (actually several different ones) for quantum
computation.

Here would not be the place to give a long technical description of
the Cirac-Zoller proposals.  However, we would like to describe an
exercise\cite{the5} which we went through subsequent to their work,
and partly motivated by it: we have tried to codify, in a
comprehensive but general way, a set of sufficient criteria that any
physical system should satisfy if it is to be suitable as a
realization of a quantum computer.  We would like to review our five
criteria, discuss briefly how Cirac and Zoller have succeeded in
meeting these criteria, and finally discuss a new proposal which we
have made\cite{LD} for implementing quantum computation in a coupled
quantum dot system, discussing how we envision the criteria could be
met in this arena.

Here, then, are the five essential criteria which we perceive for the
physical implementation of quantum computation:

1) Hilbert space control.  The available quantum states must be
precisely enumerated, and it must be known how to confine the state
vector of the quantum system to this part of Hilbert space.  In
addition, the Hilbert space should be extendable, preferably with a
simple tensor-product structure, by adding particles to the system.
For example, $n$ spin-1/2 particles have a simple spin Hilbert space
of $2^n$ dimensions.

2) State preparation.  Within this Hilbert space, it must be possible
to set the state vector initially to a simple fiducial starting state.
A simple example of this, in the spin system, would be to set all the
spins in the spin-down state.  Frequently this only requires being
able to bring the system to sufficiently low temperature that it is in
its ground state.  This is more difficult in some examples than in
others.

3) Low decoherence.  This is the criterion most closely tied to the
topic of the first part of this paper.  The coupling to the
environment (i.e., to all the rest of the Hilbert space of the world)
should be sufficiently weak that quantum interference in the
computational Hilbert space is not spoiled.  Given our current
understanding of error correction and fault tolerant quantum
computation, and given fairly benign assumptions about the nature of
the decoherence (e.g., that it acts independently on each quantum bit)
reliable computation is possible if the decoherence time exceeds the
switching time by $10^6$\cite{Pres}.  More clever fault-tolerant
techniques\cite{Kit} may well succeed in making this rather demanding
threshold number more relaxed in the future.

4) Controlled unitary transformations.  This is the fairly obvious
central requirement of quantum computing: it must be possible to
subject the computational quantum system to a controlled sequence of
precisely defined unitary transformations.  The precision requirements
are closely related to the decoherence threshold; imprecision of
unitary operations is a form of decoherence.  For convenience of
programming, it is very desirable that the elementary unitary
transformations be implementable as discrete one- and two-qubit
operations.

5) State-specific quantum measurements.  The readout of a quantum
computation, which would consist of some ordinary bit string, is to be
the result of a sequence of quantum measurements performed on the
computational quantum system.  It is very desirable (although {\em
not} necessary) that these measurements be the textbook projection
measurements of individual quanta.  It is essential that these
measurements can be made on specific, identified subsystems of the
computational state; in the simplest case, this means that it should
be possible to do a projection measurement on each qubit individually.
Recent work in nuclear-magnetic-resonance computation has shown that
certain aspects of this criterion can be relaxed\cite{Ger}: if many
identical copies of the quantum computer are available, then weaker,
ensemble measurements, rather than projection measurements, are
adequate.  It is still necessary that these ensemble measurements be
subsystem-specific, though.

The Cirac-Zoller proposal of the ion-trap quantum computer\cite{CZ}
has done a beautiful job of satisfying these criteria: 1) The Hilbert
space which they employ, the low-lying electronic and spins states of
the ions, and the quantized states of vibration of the ions in the
trap, have been thoroughly mapped out by atomic physicists in a series
of careful experiments spanning many years.  Extendibility is achieved
by adding more atoms in a line to the trap.  2) Laser cooling
techniques have enabled experimentalists to place this system in the
ground state.  (There are questions about how well this can be done
when many ions are added to the trap.)  3) Long coherence times are
well known for the internal states of the ions (although coherence
times are a bit more problematic for the vibrational states of the
ions, and the recent modifications of the proposals made by Cirac,
Zoller, and collaborators have been partly designed to circumvent this
decoherence problem).  4) Precision spectroscopic manipulation of the
ion's internal-plus-vibrational states is thoroughly demonstrated, and
a complete set of quantum logic operations is known to be achievable.
5) The availability of quantum-jump spectroscopy implies that
virtually ideal, strong, quantum-specific measurements are
available\cite{IT}.

It would appear from all this that the Landauer plea, that quantum
computing be considered at the level of real-world devices, has been
completely satisfied.  Not quite, though: despite its plausibility, the
Cirac-Zoller device is a great extrapolation --- in scale, and in
its simultaneous achievement of a variety of experimental capabilities
--- from any existing experiment in ion-trap physics, and some have
questioned whether these extrapolations, especially to very large
scale quantum computation (c. 1000 qubits) will really be possible.
There are other technical objections having to do with the fact that
the machine as envisioned does not permit parallel operations, quantum
gate operations performed simultaneously on different parts of the
device.  This is important because, from the theoretical point of
view, all the powerful results of fault-tolerant quantum computation
need this parallel capability\cite{FT,Pres}.

So, despite the brilliance of the ion-trap proposal, we have remained
motivated to propose other platforms, from very different areas of
physics, that have the potential for satisfying the five criteria for
the implementation of a quantum computer.  It is often speculated that
a solid-state physics approach will be the only plausible arena for
the massive scale-up of quantum computation which will ultimately be
desirable.  This is indeed a debatable proposition: our current ability
to hold and process untold millions of ordinary bits on a silicon chip
in no way translates into a corresponding capability to have and hold
a large number of qubits.  Still, it is a fact that there is a great
deal of basic research, projecting forth from the fabulous successes
of microelectronics, to understand the quantum properties of small
solid state devices, and it is on this fact that we hang our hopes.

We have thus been encouraged to work on a proposal for a solid-state
quantum computer, one based on quantum dots\cite{LD}.  We are sure that
the quantum behavior of such structures will be the subject of many
other articles in this volume, as the whole topic of quantum
interference in mesoscopic structures is another one which sprang
largely from the brain of Rolf Landauer.  But we think that the
proposal which we have made prescribes a much deeper use of the
quantum properties of these structures than has had been contemplated
before.

Our proposal has been outlined in detail elsewhere\cite{LD}, and we
will just give a summary here, but with a couple of added features
which have come to light recently.  We will give this outline by
discussing how we envision our five criteria for the realization of
quantum computation to be satisfied.  For pedagogical reasons, we will
visit these five items out of their order above.

1) Hilbert space: We propose using the real {\em spin} states of
electrons confined in an array of quantum dots.  Gaining control over
this Hilbert space requires, first, that the number of excess
electrons confined to each quantum dot be precisely controlled,
and in particular that the electron number be {\em odd} so that the
dot has an excess spin-1/2.  At this stage in development of
experiments in this area, many groups have succeeded in maintaining
some electron number in a dot exactly; fewer groups have the
capability of fixing the {\em absolute} number of excess electrons,
but we hope that this will be readily doable in many of the quantum
dot experiments which are envisioned.  We also require that the
electrons populate, with reasonable probability, the lowest-lying
electron orbital; in other words, we want only  spin degrees of
freedom to be available, but not  charge degrees of freedom.
This should be achievable by a combination of strong confining
potentials (i.e., small dots) and low temperatures.  It seems that for
dots substantially below 100nm in size, conventional cryogenics
(necessary for many other aspects of the proposal) should be
sufficient.

2) State preparation: Not much need be said on this point: any
conventional method of preparing the set of spins in a highly
spin-polarized state (as simple, for example, as cooling the spins in
a strong magnetic field) would be satisfactory.

5) Strong quantum measurement: It is necessary (unless we adopt the
ensemble approach introduced in nuclear-magnetic-resonance quantum
computing) to be able to measure whether the spin of any individual
dot is up or down with respect to some quantization axis.  Single-spin
measurements in the solid state are still in the future, but such
measurements have been the holy grail of quantum magnetism experiments
for many years, and we feel confident that eventually such a
measurement will be achieved by some means.  We might highlight one
suggestion which we have made\cite{LD} for how to do this which
integrates well with the technology of single-electron quantum-dot
experiments.  Suppose a tunneling barrier could be introduced into the
system whose barrier potential is spin dependent; such barriers are
well known in some areas of magnetic physics, although it has not yet
been contemplated how to incorporate them into the processing used to
create quantum dots.  The gating of such a barrier between two quantum
dots, one containing the spin state to be measured, and the other
containing no excess electron, could, at some desired instant in time,
make it possible for the electron to tunnel through the barrier only if
it is in one of the two spin states.  Then, the presence or absence of
the excess electron in the second dot, which can be done by well
understood and perfected single-electron electrometry techniques,
would serve as the desired measurement of whether the electron had
been in the spin up or spin down state.  If this technique turns out
to be infeasible, we are confident that experimentalists will use
their ingenuity to solve this problem in a much more practical way
than we can ever envision.

4) Gate operations: This is at the heart of our quantum dot proposal;
we discuss a few recent further advances in our thinking on this in
Sec. IV.  We envision a variety of mechanisms for effecting one-qubit
and two-qubit gate operations on the spin qubits of the quantum dots.
Our proposals begin with the recent development of the experimental
capability to controllably couple or decouple the states of
neighboring dots by externally controlling an electric potential
barrier between them\cite{Liv}.  In the present experiments, this
capability is used only to demonstrate that the dots can go from a
regime where a single added electron enters one of the two dots (the
decoupled situation) to a regime in which an added electron goes into
a delocalized state of the two dots (the coupled situation).  We
propose using this capability in a more subtle way: it is well known
that virtual tunneling of electrons between two spin-degenerate sites
leads to an effective exchange coupling between the spins of the two
electrons.  By turning on tunneling (by lowering the potential
barrier) for a controlled length of time, a specific two-qubit gate
operation could be achieved.  The exchange interaction leads to a
quantum gate of the ``swap'' type; for a particular duration of the
interaction (or any odd multiple of the fundamental duration), the
exchange is complete and gate is just a complete swapping or
interchange of the two spin states.  This does not constitute a very
useful two-qubit quantum gate; but if the interaction is left on for
half of this fundamental swapping time, the resulting ``square-root of
swap'' operation, in conjunction with other gates which we will
discuss next, would provide an efficient basis for programming any
desired quantum computation.

Square-root-of-swap is still not as powerful a quantum gate as is
needed theoretically, because it respects rotational symmetry in spin
space.  Thus, it leaves the total angular momentum of the spin system,
and its projection on any quantization axis, unchanged.  But in
quantum computation it is desirable to be able to rotate the state
vector from any state to any other state.  For this reason, it is very
desirable to supplement the two-bit swap-type gate with other gates
which break the spin-space rotational symmetry.  This is very easy to
do from a theoretical point of view, it just requires adding a simple
family of one-qubit gates.  Unfortunately, the experimental
implementation of such gates is surprisingly problematic, as it
involves the application of magnetic fields locally to an individual
spin (and not to the surrounding ones).  This is a daunting technical
requirement, which, we readily admit, would require quite heroic
experimental efforts to achieve; we hope that there would be other
ways to achieve this which we are still investigating, perhaps if it
were possible to perform local electron-spin-resonance operations on
the system.

But for the time being, we offer a few tentative ideas for how this
application of a localized magnetic field might be conceivable, with
apologies for not having been able to see how to make it any simpler.
Since, in the measurement scheme which we have proposed above, we have
suggested the incorporation of magnetic materials into the system (to
make the spin-dependent tunneling barrier), we could envision using
such materials to accomplish the one-bit gates.  For example, if a
piece of such a magnetized material were placed near the quantum dot,
such that by lowering a gate potential, the electron could be made to,
at some desired time, partially penetrate the magnetized barrier, the
electron spin would precess around this internal magnetization and the
one-bit operation could be achieved.  A magnetized dot which the spin
state could be swapped into and out of could have the same effect.  If
magnetic materials were undesirable, one could envision various local
coil arrangements or magnetized probe tips which could also give the
desired one-bit operations.  Hopefully, more ingenuity will lead to
more elegant solutions to this problem.

3) Coherence times: Consideration of this criterion for quantum
computation also leads us into speculative territory, but one which we
are reasonably hopeful about.  In the usual mesoscopic experimental
regimes, it was rare to find decoherence times even as long as one
nanosecond; in mesoscopics, however, it was always the decoherence of
a {\em charge} degree of freedom which was being studied.  There is
every reason to believe, from a theoretical point of view, that the
coherence of electron {\em spin} states should, under favorable
circumstances, be much longer.  Generally speaking, the coupling of
the environment to spin is weaker than to charge.  There is as yet
very little experimental indication of how long these spin coherence
times could be.  Kikkawa {\em et al.}\cite{Kik} have observed free
induction decay for a population of photoexcited electron spins in a
quantum well.  The $T_2$ measured in this decay, which is a lower
limit on the decoherence time for the spins, was seen to be several
nanoseconds.  We may in addition consider the decoherence time for the
spins of itinerant electrons in a 2D quantum well to be a lower limit
on the time for electrons in a similar material but confined to a
zero-dimensional structure.  For these reasons, we believe that the
Kikkawa observations should be just considered a very early starting
point in the search for long spin coherences, and that increases in
these times of many orders of magnitude would not be out of the
question.  It would indeed be extremely desirable to find a system
with a decoherence time of, say $10^{-3}$ sec., since the speed of the
desired gate operations would be scaled to this time; in any
foreseeable experiment it would be very interesting to make the gate
times a few orders of magnitude faster than the decoherence time.
This would mean gate operations going at a MHz rate, which would we
think be a fairly comfortable regime for AC manipulations of low
temperature electronic systems.

We hope very much to engage in a dialog, in the Landauer style, with
experimentalists and other interested parties to improve this quantum
dot proposal through critical discussion.  We are certain that the
solutions which we have proposed for satisfying the criteria of quantum
computation are not optimal, and perhaps on further examination they
will prove to be laughable; but we cannot see any ``show stoppers'' at
this point, and we remain optimistic that solid state quantum
computation
will indeed be possible and will indeed be a very exciting line for
fundamental experiments in quantum physics.

\section{Recent results on coupled quantum dots}

To obtain a more quantitative understanding of the origin of the
exchange coupling occurring in the effective two-spin Heisenberg model
and to determine its magnetic and electric field dependence, we have
begun recently\cite{BL} to investigate coupled quantum dots from a
more microscopic viewpoint. In the following we wish to report on
these preliminary findings.

Our investigations have been motivated by recent advances in the
physics of semiconductor quantum dots that were fabricated in a 2DEG
GaAs system by Tarucha et al.  \cite{Tarucha} These experimentalists
have demonstrated that such dots are well-described by a parabolic
confinement potential (of energy $\hbar \omega=3$meV ) and that one can
fill in one electron after the other (starting with an empty dot) in a
controlled way.

Armed with this information it is now reasonable to expect that it
should be within experimental reach (as envisioned in point (4) of
Sec. III) to couple two such dots (containing only a few electrons)
via a tunable or non-tunable barrier (as has already been achieved in
bigger dots \cite{Liv}).  The physics of such a system can then be
described by adopting the lines of reasoning used in molecular theory.
To put it in other words, in the same way as one can consider an
isolated quantum dot as an artificial atom that obeys e.g. analogs of
Hund's rule when electrons are added to the shells, one can now
consider the problem of coupled quantum dots as the problem of
artificial molecules, or more generally as the problem of ``quantum
dot chemistry."  Like in ordinary chemistry, we can use techniques
such as the Heitler-London method or more refined approaches such as
the Hund-Mullikan ansatz, hybridization etc.  to obtain the low-lying
energies. One of the main differences between ordinary atoms and
quantum dots is that the attractive forces between nuclei and
electrons are now replaced by the parabolic confining potential that
can be controlled externally by changing the gate voltage. The
associated Bohr radius $a_B=\sqrt{\hbar/m\omega}$ ($m$ is the
effective electron mass) is typically in the range of tens of
nanometers and thus much larger than in real atoms. One important
consequence of this is that (coupled) quantum dots are much more
sensitive to external magnetic and electric fields.  As we will see
below it is this field sensitivity which allows one to tune the
exchange ``constant" to zero as a function of uniformly applied
external fields, the strengths of which are easily accessible in
standard set-ups.

To be specific let us consider the simplest case, namely two circular
quantum dots of radius $a$ lying in the same plane and whose centers
are separated by 2$a$. Each dot contains one electron of spin 1/2
which interact via the (possibly screened) Coulomb interaction.  It is
then straightforward to write down an explicit Hamiltonian that
captures the physics just described and that will allow us to perform
some more concrete evaluations:
\begin{equation}
H=\sum_i h_i + \sum_{i<j}v_{ij},
\end{equation}
where
\begin{eqnarray}
h_i&=&{1\over{2m}} ({\bf p}_i-{e\over{c}}{\bf A}({\bf r}_i))^2 +ex_iE
+{m\omega^2\over 2}[{1\over 4 a^2}(x_i^2-a_i^2)^2+ y_i^2]
+g\mu_B {\bf S}_i\cdot{\bf B}/\hbar
\,\, ,\nonumber \\ v_{ij}&=&{e^2\over{|{\bf r}_i-{\bf r}_j|}}\,e^{-\mu
|{\bf r}_i-{\bf r}_j|}.
\end{eqnarray}
The gauge potential ${\bf A}({\bf r})=(-yB/2,xB/2,0)$ describes the
effect of the constant magnetic field ${\bf B}=(0,0,B)$, and ${\bf
E}=(E,0,0)$ is some electric field applied along the $x$-axis
connecting the dots. These dots are located at {\bf a}$_i=(a_i,0,0)$.
The coupling of the dots is described in terms of an $x^4$-potential
with $\hbar\omega$ given by the parabolic confining energy of a single
isolated dot. The change of barrier height between the dots can then
be described by changing the interdot distance $|a_1-a_2|=2a$. The
last term in $h_i$ is the Zeeman term.  The Coulomb interaction is
described by $v_{ij}$ with $\mu$ being some effective screening
parameter. The motion of the electrons is assumed to be planar, i.e.
${\bf r}=(x,y,0)$.

This Hamiltonian cannot be solved exactly but we can make progress
with the help of variational (or numerical) techniques to find for
instance the exchange constant given by the difference between singlet
and triplet energies.  In particular, in the Heitler-London
approximation and making use of the Darwin-Fock solution for the
isolated dots we find (omitting all details of the calculation
\cite{BL}),
\begin{equation}
J={\hbar\omega \over
{\sinh\left(2bd^2\frac{2b^2-1}{b^2}\right)}}\,
\left[c\sqrt{b}\,\left(e^{-bd^2}\,I_0\left(bd^2\right)-e^{bx\frac{b^2-1}{b^2}}\,I_0\left(bd^2\frac{b^2-1}{b^2}\right)\right)
+{3\over 4 b}\left(1+bd^2\right)\right]\,\, ,
\label{exchange}
\end{equation}
where $c=\sqrt{2\pi}(e^2/a_B)/2\hbar\omega$ is the ratio of Coulomb
energy to confinement energy, and $I_0$ the zeroth-order Bessel
function. For $\hbar\omega=3$ meV \cite{Tarucha} we
have $a_B=19$ nm and thus $c=2.4$. Further, the parameter
$b=\sqrt{1+\omega_L^2/\omega^2}$, with $\omega_L$ being the Larmor
frequency, describes the effect of the orbital diamagnetism, it
becomes appreciable in the Tesla range since $\omega_L/\omega=0.3$
(B/T). The dimensionless distance between the dots is given by
$2d=2a/a_B$. For the moment we have set $E=0$ and assumed a bare Coulomb
interaction (which is a reasonable assumption for the two--electron
system). Note that the energy scale of the exchange coupling is given
by the confinement energy $\hbar \omega$.

A plot of $J/\hbar\omega$ versus magnetic field is given in Fig. 1
(for $d\approx 0.7$).  The most interesting feature of the
Heitler-London
result is the fact that
due to the influence of the orbital diamagnetism the exchange $J$
passes through zero (at a field value of about 1 T) and thus changes
from antiferromagnetic ($J>0$) to ferromagnetic ($J<0$) coupling.
This suggests  again a
novel mechanism with which one can tune the exchange coupling J to
zero. Of course, $J$ can also be tuned to zero asymptotically.
We stress that the magnetic field is not local but extends
uniformly over the two dots, and such a uniform field can be easily
accounted for in the XOR operation.
Finally we note that for vanishing B-field and for c values
with $c<2.8$ (which is in the range of experimental interest)
J is positive for all distances $d$ (i.e. the singlet
state is lower in energy than the triplet state),
as it must of course  be the case
for a two-electron system on general grounds. The Heitler-London
approximation
for $J$ breaks down (i.e. $J$ becomes negative even for $B=0$)
for certain $d$'s when $c$ exceeds 2.8.

Next, adding an electric field $E$ will lead to a simple shift of
$J$,
\begin{equation}
J_E=J+{3\hbar\omega/2\over
d^2\sinh\left(2bd^2\frac{2b^2-1}{b^2}\right)}\left({eEa\over\hbar\omega}
\right)^2\,\, .
\end{equation}
This expression is valid for not too large electric fields with
$eEa\over\hbar\omega$ not exceeding one.  Thus, within the
Heitler-London approximation we find that such a field (or
biasing voltage) can then be used to also tune the exchange
constant. Both of these tuning mechanisms could be used alternatively to
or in conjunction with a gate between the dots by which one can tune
the barrier height by varying the gate voltage (in our model, this
tuning mechanism can be accounted for by varying the dot distance
$d$).

It is interesting to examine the magnetization $M$ (along the
$z$-axis) of the coupled dots, as this quantity can give independent
information about intrinsic parameters such as the exchange coupling
but also about the interplay between orbital diamagnetism and spin
paramagnetism. This quantity can also be calculated in the
Heitler-London approximation\cite{BL}. However, we shall not write
down the lengthy expression here and content ourselves with a plot of
the magnetization versus B-field, see Fig. 2, where we stay in the
low temperature regime of Ref.\cite{Tarucha} with $\hbar\omega=170$
k$_B$T, which corresponds to an electron temperature of T$=200$mK.  The
striking feature to be noted here is the initial diamagnetic response
(with the spins being antiparallel) followed by a sudden jump at about 1
Tesla. Indeed, this jump can be traced back to the
change of sign in the exchange constant.  After the jump, the response
becomes diamagnetic again (with the spins being now parallel)
and finally approaches saturation
asymptotically. Thus the Heitler-London approximation
suggests that the sudden switch around 1 Tesla allows one to get
direct information about the exchange constant from
the magnetization.

It does not need to be stressed of course that it
will be rather difficult to measure the magnetization of only two
electrons, as the magnetization is only of the order of a few Bohr
magnetons.  Still, in a first set of experiments one can again envision
(as in \cite{LD}) a scaling up to many independent systems of two
coupled quantum dots. Also, the present status of cantilever
technology is capable of measuring magnetic moments on the order of
a single Bohr magneton! It would be interesting to explore the
possibility whether one could use magnetic force microscopes etc. to
measure such magnetization effects.

The above analysis can (and should !) be refined by making use of
the Hund-Mullikan
(or LCAO) method and by including sp-type of hybridization effects
(which however are balanced by orbital field effects).  These
calculations become rather involved \cite{BL},
and we will report on them elsewhere. In
principle, it is possible to solve this problem to arbitrary accuracy
by making use of powerful numerical techniques developed in molecular
physics. [It is amusing to mention parenthetically that in these
numerics one approximates the atomic wave functions by Gaussians
mainly for technical reasons; here in our case of quantum dots with
parabolic confinement this would in fact be exact and a much better
convergence can be expected.]

It is worth mentioning that spin-orbit effects can essentially be
neglected in the case of only very few electrons per dot. Indeed, the
spin-orbit interaction in a quantum dot with parabolic confinement
takes the form $H_{so}={\omega^2\over m c^2} {\bf L}\cdot {\bf S}$,
where ${\bf L}$ is the angular momentum of the confined
electron which is of the order of $\hbar$. Thus
we can estimate that $H_{so}/ \hbar\omega\sim \hbar\omega/ mc^2\sim
10^{-7}$ for above values and with an effective electron mass found in
GaAs.

It is clear by now that the above analysis can be extended to
situations with more than one electron per dot, although the complexity
of the problem increases rapidly.  We hope to report soon on our
progress in this direction.

Finally, it is a most rewarding aspect of this area of research, which
we have the privilege to bring to Rolf Landauer's attention, that even
apart from our ultimate goal of building a working quantum computer
there is plenty of fascinating and novel physics to be discovered on
our way that will keep us (and hopefully our experimental colleagues !)
quite busy for a while.

\acknowledgments
We would like to thank G. Burkard for useful discussions.

\begin{figure}
\epsfxsize=15cm
\leavevmode
\epsfbox{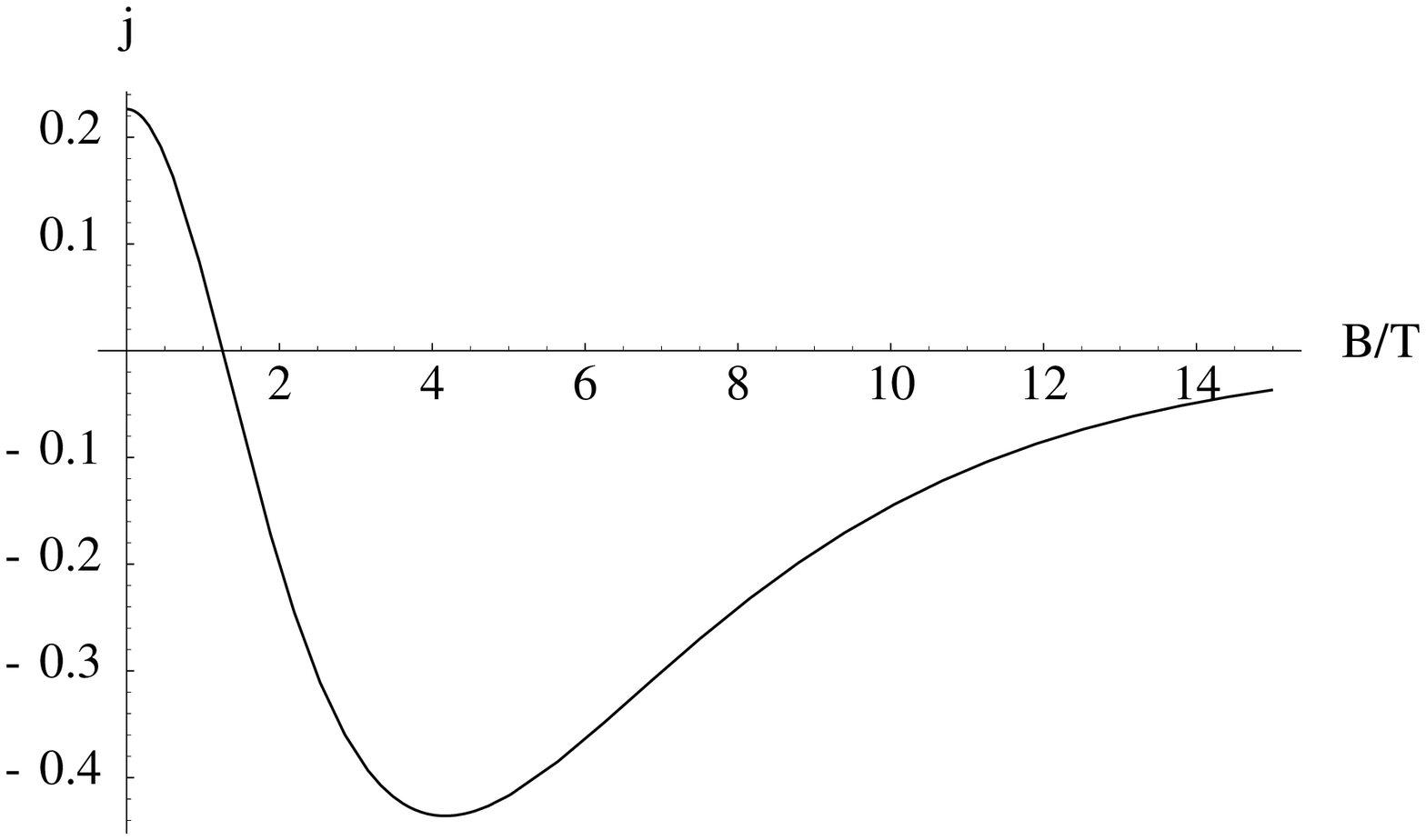}
\caption{Exchange coupling $j=J/\hbar\omega$ versus magnetic field $B$
in units of Tesla as calculated within the Heitler-London
approximation, Eq. (\protect\ref{exchange}). The ratio of Coulomb
energy to confinement energy is $c=2.42$, and the the dimensionless
interdot distance $d=a/a_B$ is set to 0.7. For interpretation see
main text.}
\label{figure1}

\end{figure}

\begin{figure}
\epsfxsize=15cm
\leavevmode
\epsfbox{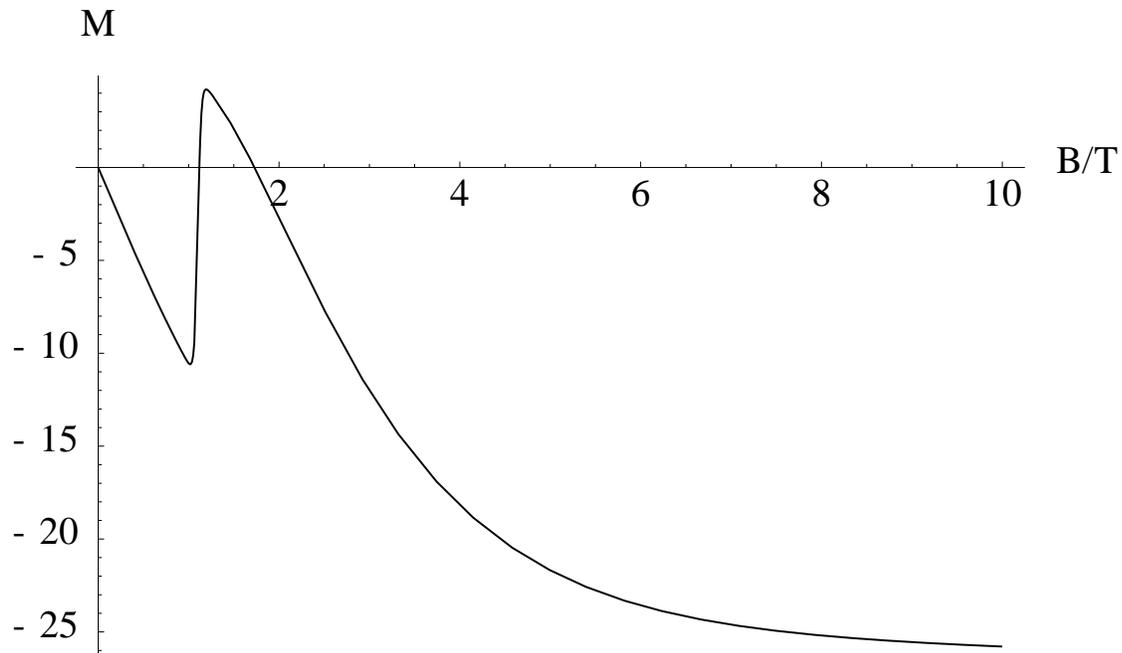}
\caption{Magnetization ${\bar M}=M \mu_B$ versus $B$-field in units of
Tesla, as calculated within the Heitler-London approximation.
Here, $\mu_B$ denotes the Bohr magneton. Note that the maximum amplitude
of ${\bar M}$ is about
$25 \mu_B$ ($c$ and $d$ are the same as in Fig. 1). For further
interpretation see the main text.}
\label{figure2}

\end{figure}
\end{document}